\documentclass[12pt,journal,onecolumn]{IEEEtran}
\ifCLASSINFOpdf
\else
\fi
\usepackage[utf8]{inputenc} 
\usepackage[T1]{fontenc}    
\usepackage{booktabs,tabulary}
\usepackage{multirow}
\usepackage{adjustbox}
\usepackage{fancyhdr}
\usepackage{kantlipsum}
\usepackage{hyperref} 
\usepackage{rotating}
\usepackage{amsmath}
\usepackage{color,soul}
\usepackage[verbose=true,a4paper,margin=1in]{geometry}
\usepackage{lipsum}
\usepackage{setspace}
\usepackage{wrapfig}
\usepackage{float}

\hypersetup{ 
	colorlinks=true,
	linkcolor=black,
	filecolor=black,      
	urlcolor=black,
	citecolor=black,
}

\DeclareRobustCommand*{\IEEEauthorrefmark}[1]{%
	\raisebox{0pt}[0pt][0pt]{\textsuperscript{\footnotesize #1}}%
}

\begin{document}

\begin{titlepage}
	
	\vspace{2cm}
	
	\begin{center}
		\Huge
		EEA Conference \& Exhibition 2021, \\
		30 June - 1 July, Wellington, New Zealand
		
		\vspace{1cm}
		\Large
		\textbf{Monitoring and Detection of Low-current High-Impedance Faults in Distribution Networks}
	\end{center}

	\vspace{2cm}
	
	\begin{center}
		Anwarul Islam Sifat\IEEEauthorrefmark{1}, Fiona J. Stevens McFadden\IEEEauthorrefmark{1}, Ramesh Rayudu\IEEEauthorrefmark{2}, Joseph Bailey\IEEEauthorrefmark{1}
	\end{center}

	\begin{center}
		\IEEEauthorrefmark{1} Robinson Research Institute, Victoria University of Wellington
		
		\IEEEauthorrefmark{2} School of Engineering and Computer Science, Victoria University of Wellington
	\end{center}

	\vspace{2cm}
	
	\begin{center}
		Presenter: \\
		
		Anwarul Islam Sifat\\
		
		Fiona J. Stevens McFadden\\
		
		Robinson Research Institute, Victoria University of Wellington
	\end{center}
	
\end{titlepage}

\section*{Abstract}
Faults in electricity distribution networks have the potential to ignite fires, cause electrocution, and/or damage the system itself. High-current Low-Impedance Faults (LIF) are typically detected and mitigated via over-current, distance, directional relays, fuses, etc. In contrast, while High Impedance Faults (HIF) are equally hazardous, they are much more challenging to detect due to the fault current being much lower than load currents and their time-varying and nonlinear behaviour. A solution for HIF detection will require, firstly, the continuous observation of system current waveforms and, secondly, the development of sophisticated pattern recognition algorithms to apply to these measurements. However, New Zealand distribution networks are extensive and largely unmonitored beyond the substation, and suitable HIF detection schemes are still an ongoing research challenge. 

Robinson Research Institute has been developing a solution to enable widespread monitoring of overhead lines in distribution networks and detecting faults. The sensing system we have developed and are evaluating utilizes magnetic field sensors (specifically the Giant Magneto Resistive type), pole-mounted at a distance from the overhead lines for ease of installation. These sensors are an attractive option compared to traditional current transformers (CTs) due to their non-contact sensing ability, wide frequency bandwidth, low cost, miniature size, and ease of integration with required digital componentry. Coupled with the algorithms we have developed, these sensors allow observation of the individual overhead line currents and detect both Low Impedance and High Impedance faults.  

To date, we have built a physical test facility for power system fault analysis and the development and evaluation of our sensing and fault detection system. We have simulated LIF and HIF with different fault surface materials and load-switching events. From the data collected, we have characterized the unique fault behaviour for both LIF and HIF in 400V networks and trained a Deep Learning classifier to recognize the type of fault present from its unique signature. We have developed an outdoor pole-mountable sensing system and have installed this in Wellington Electricity’s network for ongoing data collection and evaluation.  

This paper will describe the test facility and our experience developing and implementing the sensing system. The widest range of HIF phenomena observed was in the fault experiments involving the tree branch. For brevity, therefore, this paper reports on the results of just these tree-branch experiments. HIF faults on other surface materials will be reported elsewhere. Finally, we will detail the pole-mountable sensing system installed in Wellington Electricity’s network and the outcomes thus far.


%
\IEEEpeerreviewmaketitle

\pagebreak
\section{Introduction}
Detection of power system faults is an ongoing challenge in the power distribution industry. Faults overstress the existing infrastructure and push operation close to its stability limit. In addition, faults can endanger human life \cite{abadee}. Broadly, faults can be classified as low (LIF) or high impedance faults (HIF). Due to the resulting high current, LIFs are easily detectable by conventional protection schemes, e.g. circuit breakers (CB) and protective relays \cite{fernandez2001impedance}. HIFs are when the impedance of the fault path is high and stochastic in nature. These generate a much lower fault current that cannot be easily recognized or reliably detected by conventional relays. A reliable and generalized protection scheme is required that recognizes the patterns of all types of fault.

The New Zealand distribution network is extensive, with  about 98,549 km of overhead lines, and are sparsely monitored despite them being an economically significant part of the power networks \cite{edb2020a}. Of these, 24,321 km are low voltage (<1kV) distribution networks (LVDN) that have no monitoring equipment at all. The networks are susceptible to unplanned outages. An average of 130 minutes of unplanned interruptions per customer was recorded during 2008-2020 period across New Zealand's distribution networks, which cost distribution companies around NZD \$552,000/min \cite{edb2020}. In addition, these unplanned outages incur a less easy to measure but significant cost to customers (the cost of lost load to customers has been found to be twice that of a planned interruption).

CTs have been the prevalent device  in distribution systems for monitoring. However, a CT has innate nonlinear magnetic core characteristics, high core losses, narrow bandwidth, and is incapable of DC measurements. All generations of CTs use a metal core, which makes them heavy and places an additional burden on the overhead lines’ mechanical load. Also, the core saturates under high-frequency (HF) transients \cite{ferreira2010noninvasive}, causing clipped-out instantaneous current measurement. For these reasons plus their cost, the widespread use of CTs in distribution networks as a solution for the lack of monitoring is not attractive.

As an alternative, magnetic sensors, particularly Giant Magneto-Resistive (GMR) sensors \cite{Baibich1988}, are an attractive option due to their non-contact sensing ability, low cost, miniature size, and wide frequency bandwidth \cite{ziegler2009current}. A GMR sensor implemented in the vicinity of overhead power lines can capture the alternating magnetic field generated by the flow of current.

In this paper, we describe a purpose-built physical simulation test facility that we have used to test the dynamic response of GMR sensors under fault conditions and normal system events. The characteristics of tree branch (TB) HIF from both current and magnetic measurements, are summarized and unique features are observed. In addition to our focus on fault detection we have investigated the use of GMR sensor measurements for overhead lines current sensing and this evaluation is also presented. Finally, we detail a protype pole mounted sensing system that we have built and installed in Wellington Electricity’s network.

\section{Description of Robinson Research Test Facility}
We have designed and implemented an indoor 400 V network test facility at Robinson Research Institute. Its main intended use was to generate and characterize the typical power system faults using current and magnetic field measurements. The behaviour of a fault depends on the power system configuration, its voltage-frequency level, and types of connected loads, and therefore can only be fully realized through practical experiments. A controllable and safe way to do such practical experiments is in such a purpose-built power system test facility. 

In addition to the use of this test facility for fault simulation studies, it also has the capability to be used to assess other electrical and electronic equipment. For example, the modular design of the test facility could allow a user to characterize miniature circuit breakers (MCCB) under fault conditions, evaluate conventional and new current sensors or perform micro-grid simulations.

\begin{figure}
	\vspace{-1em}
	\centering
	\includegraphics[width=1\textwidth]{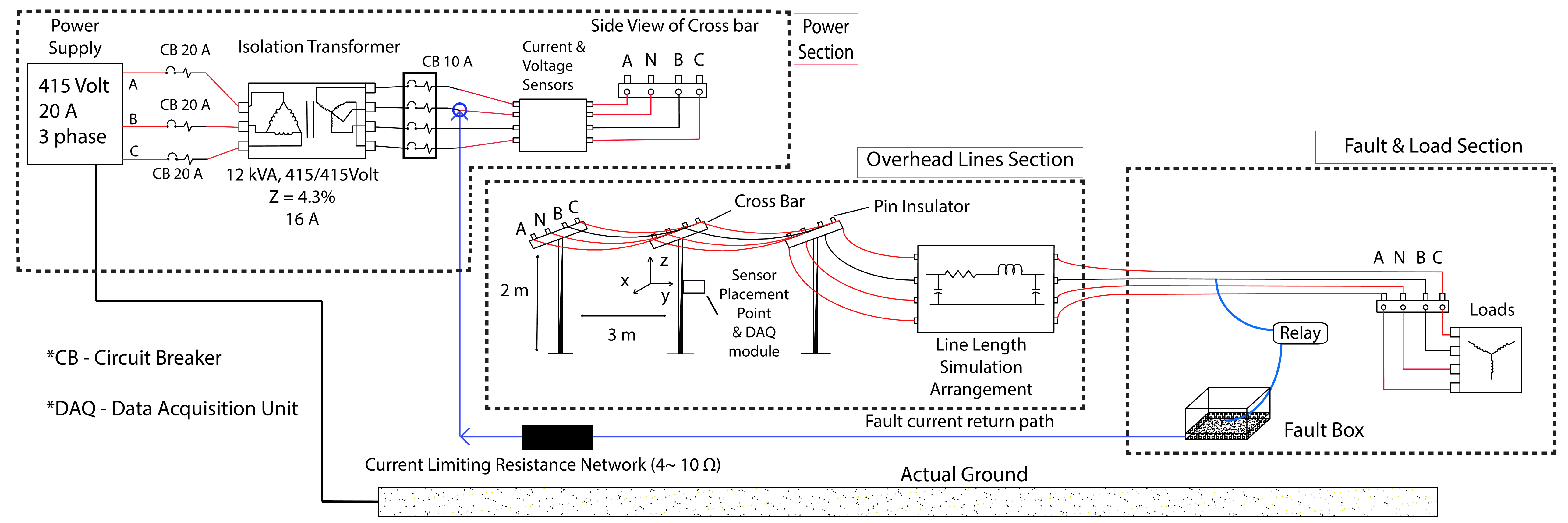}
	\caption[Test Facility Schematic Diagram]{Test facility schematic 
		diagram. The blue line indicates the fault current return path (FCRP) 
		to the neutral point of the transformer.}
	\label{fig_plan_outline}
	\vspace{-1em}
\end{figure}

A schematic of the test facility is shown in Fig. \ref{fig_plan_outline}, and key components are as follows. 
 
\subsection{Isolated Power Source}
To protect the source supply from unwanted tripping during fault experiments we use an isolation transformer. The transformer has no neutral connection in its primary side and we did not connect the secondary side neutral to the ground. All components of the facility are connected to the transformer secondary winding neutral including the fault current return path (Fig. \ref{fig_plan_outline}), meaning the secondary side is a closed loop.
\subsection{Overhead Lines Representation}
\begin{figure}
	\vspace{-1em}
	\centering
	\includegraphics[width=0.88\textwidth]{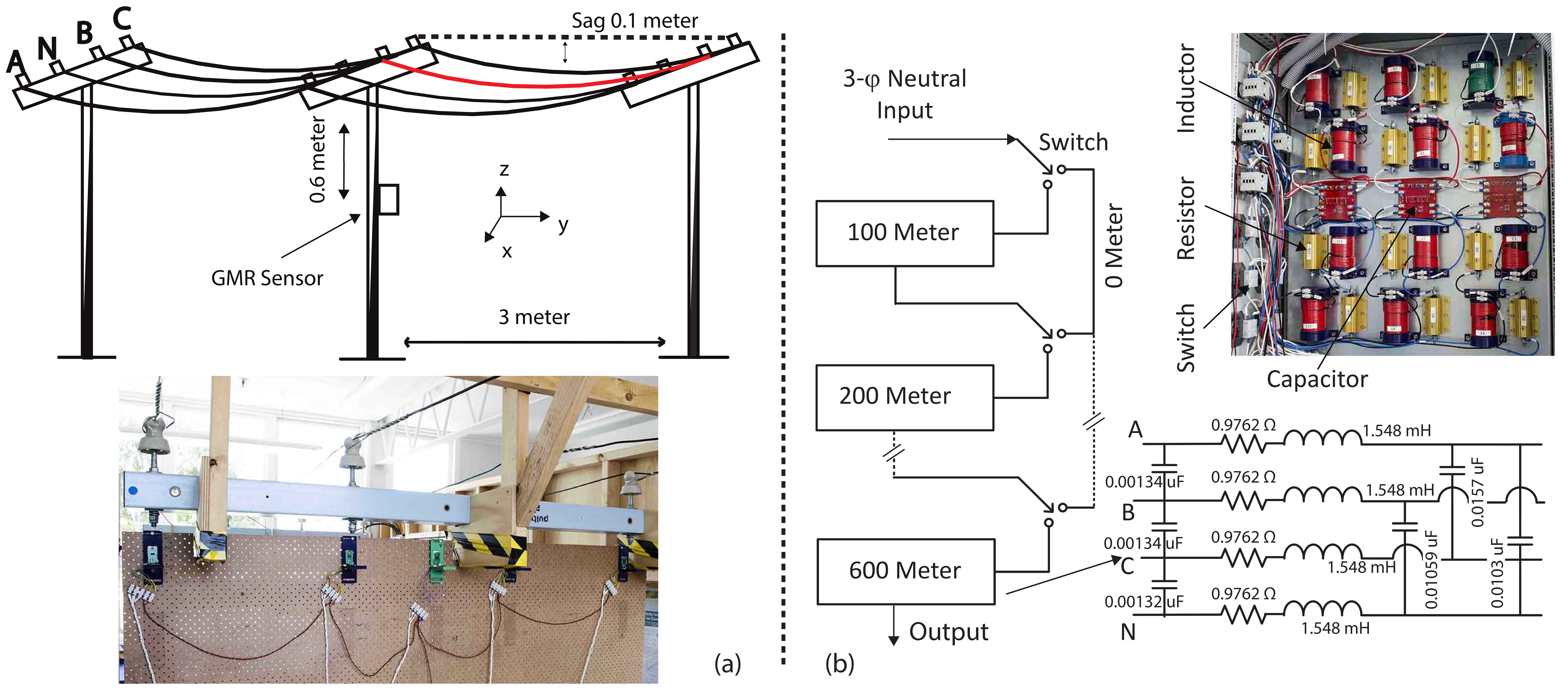}
	\caption[]{Physical representation of overhead lines. (a) Overhead Lines spatial arrangement. (b) circuit length simulator; a physical model of 600m  with a step size of 100m.}
	\label{fig_OHLS}
\end{figure}

Our facility models the typical circuit length of up to 600m for a low voltage (400 V) distribution network. The total circuit length is the combination of:

\begin{itemize}
	\item An actual spatial representation of overhead lines.  Having a spatially correct overhead line section was important to allow collection of data for the development of a contact-less magnetic sensor unit for monitoring distribution network overhead lines and detecting faults. In contrast to the typical span length of 30m we use two 3m spans, as 80\% of the total magnetic field is generated within the first 3m of the line \cite{Sifat2019}.
	
	\item The remainder of the circuit length was physically implemented using a pi model concept, which uses lumped resistance, inductance, and capacitance (Fig. \ref{fig_OHLS} (b)). The modified Carson's equation \cite{Kersting2012} was used to estimate the overhead distribution line’s parameters (i.e., series impedance and mutual capacitance). There are six pi model modules, each 100m to allow circuit lengths of 100 to 600m of lines. A user can manually switch into any circuit length using external switches.
\end{itemize}

\subsection{Fault and Load Simulator:}
The fault and load section includes components to simulate low and high impedance single line to ground (SLG) faults under unbalanced load conditions. Both broken and unbroken conductor faults can be simulated.

\begin{figure}
	\centering
	\includegraphics[width=0.88\textwidth]{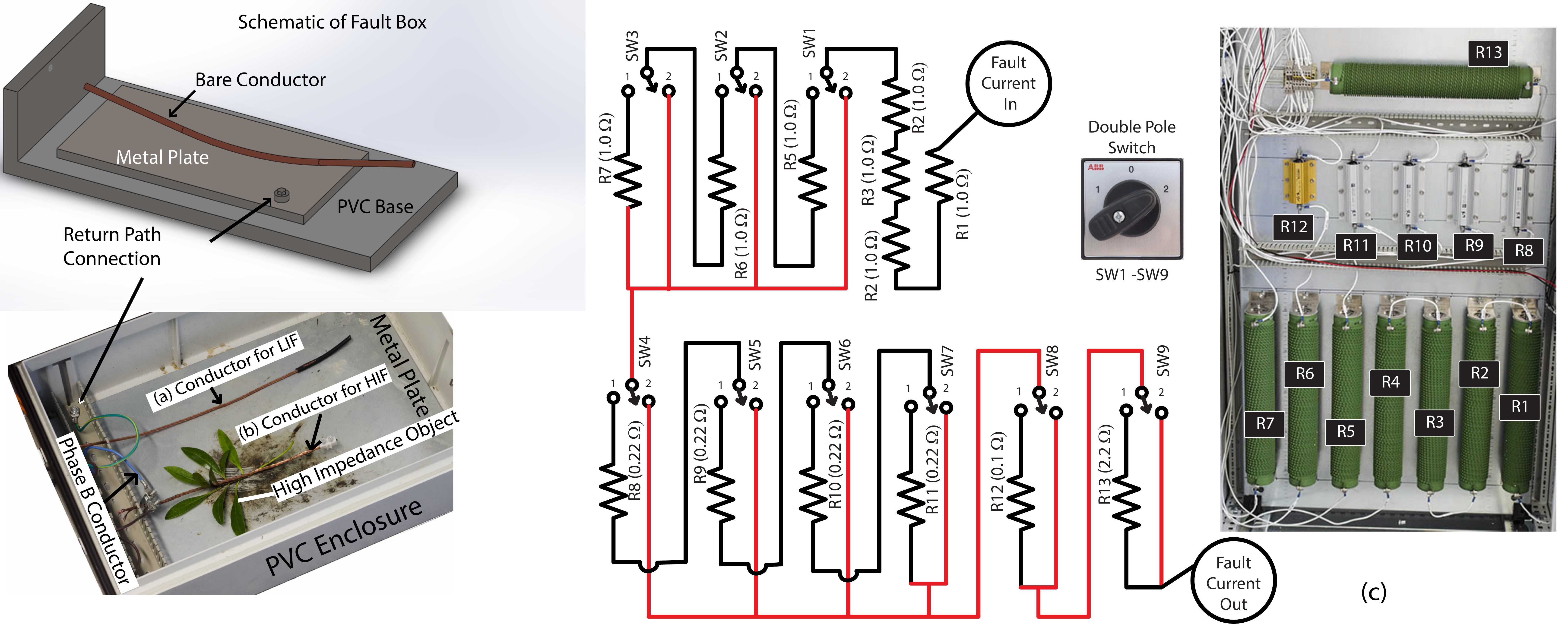}
	\caption[Staged Fault Apparatus]{Fault staging apparatus (fault box) showing setup of (a) LIF; (b) HIF with a tree branch used as a high impedance object (bottom figure). (c) Current limiting resistor network connected in fault path connection. The network resistance can be tuned to generate a fault current of 26 to 46 A (RMS) using double pole switches.}
	\label{fig_faultbox}
	\vspace{-1em}
\end{figure}

The fault apparatus is shown in Fig. \ref{fig_faultbox}. The apparatus is a plastic box with a metal plate in the base that is connected to the transformer neutral via an arrangement of current limiting resistors (Fig. \ref{fig_faultbox} (c)). A bare conductor is connected in parallel with the phase conductor to be faulted, with a relay switch connecting these. 

\begin{figure}
	\vspace{-1em}
	\centering
	\includegraphics[width=0.85\textwidth]{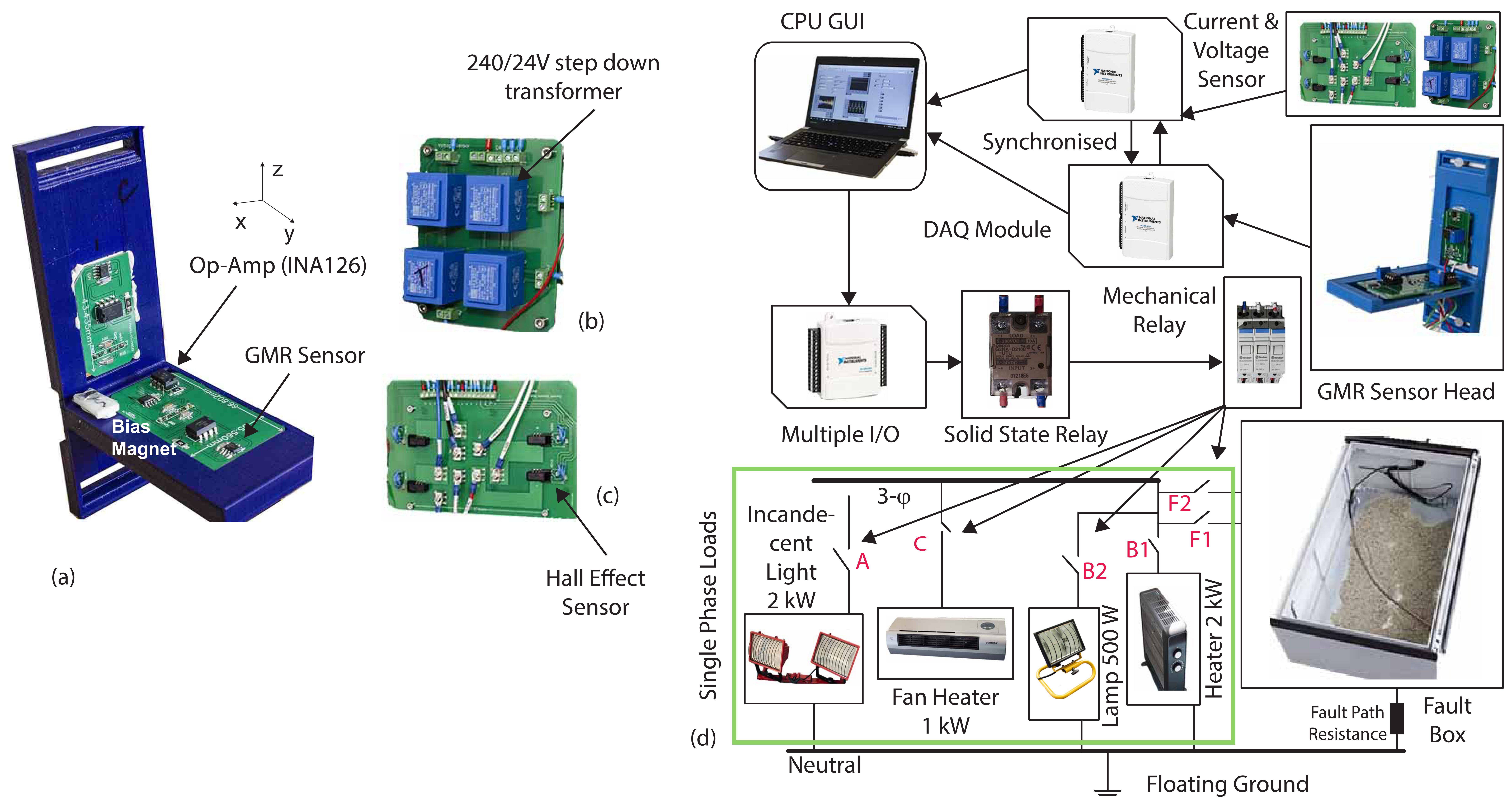}
	\caption{Sensor details and control and data acquisition system architecture. (a) The three-axis GMR (AAH002 \cite{nve}) sensor, where the sensitive axes of each sensor were directed towards each of the \textit{x}, \textit{y}, and \textit{z} directions. (b) Voltage measurement using a step down transformer incorporated with a voltage divider circuit to measure the voltage. (c) Phase and neutral current measurements with four hall effect current sensors (LEM HO 25-NP) \cite{LEM2018} mounted on a PCB circuit board trace.(d) Phase B load has two subsets, B1 and B2. The SLG fault is initiated by switching F1, and the broken conductor to the ground fault is initiated by switching F2. Both switches are connected to the fault box.}
	\label{fig_daq_}
\end{figure}

The load section presently includes two types of commonly used household loads: heaters (radiant and fan) and an outdoor light unit (Fig. \ref{fig_daq_} (d)). Therefore, the loads have a mix of resistive and inductive elements. Other loads can readily be incorporated and we are currently expanding the range of loads we can use.

Fault initiation and load switching are controlled by a computer-based (CPU) graphic user interface (GUI) via a national instrument (NI) multiple input/output (I/O) module that drives high-voltage mechanical relays through solid state relays (SSR) (Fig. \ref{fig_daq_} (d)). 
\subsection{Sensors and Data Acquisition}
The test facility incorporates voltage, current, and magnetic sensors to allow the power system faults and load events to be fully analyzed (Fig. \ref{fig_daq_}). The analog signals from these sensors were acquired synchronously using multiple data acquisition units.
\pagebreak
\section{Results: Characteristics of 400V Faults} \label{results}
In the test facility we have conducted experiments of low and high impedance faults involving tree branch and the typical 400V system event of load switching. HIFs were tested on different surface materials: sea sand, clay soil and a tree branch. In total, \textbf{270} experiments have been conducted with three different GMR sensor head positions and six circuit lengths. For brevity, representative data from the tree branch fault experiments (TB-HIF) is presented here for a GMR sensor head 0.6m below the lines and a 600-m circuit length.

\begin{figure}[]
	\centering
	\includegraphics[width=0.85\textwidth]{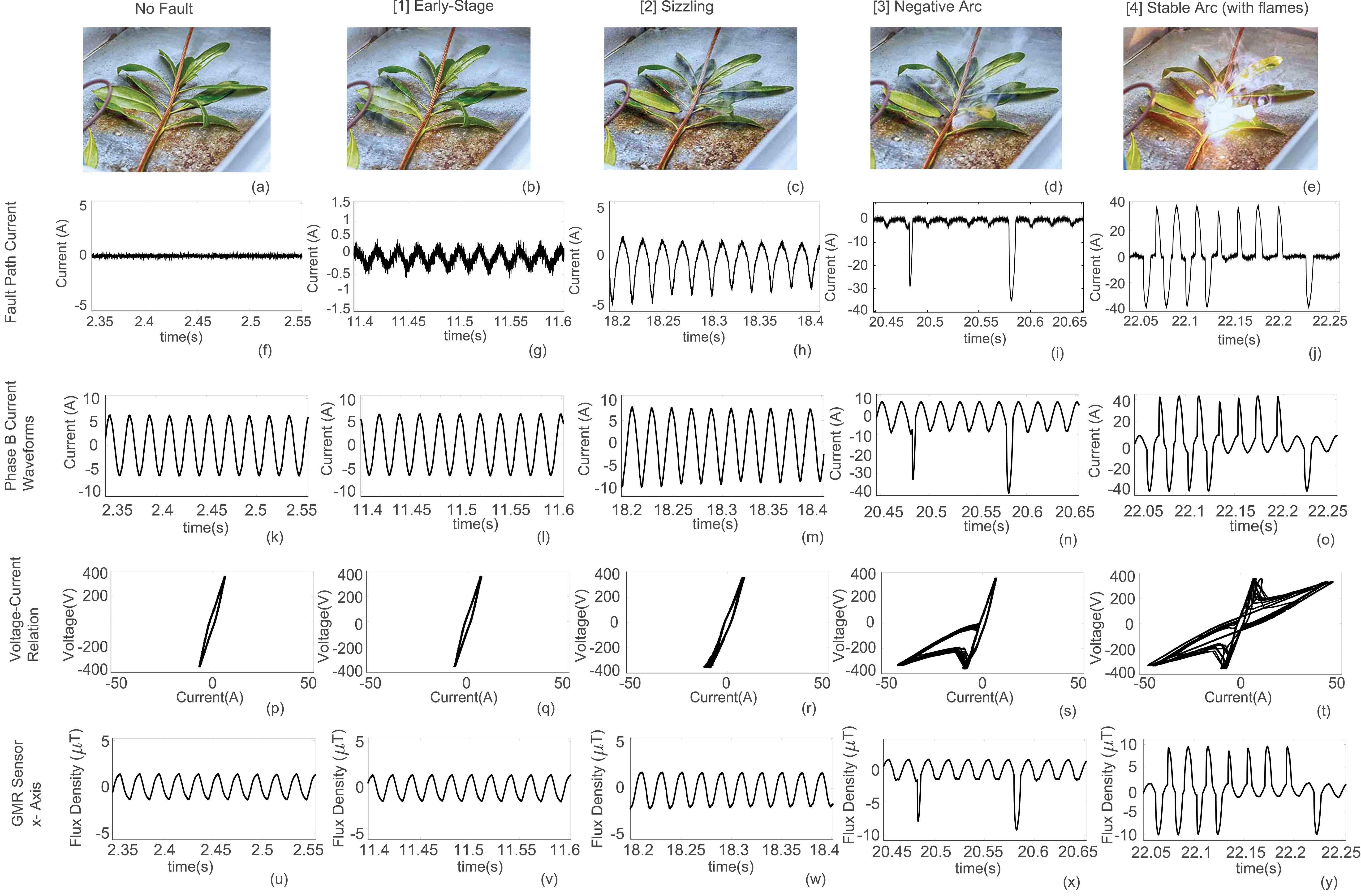}
	\caption{Snapshots on tree branch HIF stages (1st row) and measured experimental data. The stages are profiled in the fault path current, which excludes load current (2nd row), phase currents (3rd row), and magnetic flux density measurements (5th row). The phases of non-linearity between voltage-current are illustrated via \textit{v-i} curves (4th row).}
	\label{fig_TBHIF}
	\vspace{-1em}
\end{figure}

Four unique stages were observed in the TB-HIF experimental data (Fig. \ref{fig_TBHIF}):

\begin{itemize}
\item Low current initial stage, where the fault current is a saw-tooth wave-shape (Fig. \ref{fig_TBHIF} (g)) and much lower than the load current (Fig. \ref{fig_TBHIF} (l)). 

\item Sizzling stage, which has a persistent audible humming noise with visible sparks. The fault current signal looks like a full-wave rectified wave-shape (Fig. \ref{fig_TBHIF} (h)).

\item Negative half cycle arcing, where the arcs occur intermittently and only on the negative half cycle. The arcing produces a crackling sound and burns the tree branch. A significant distortion in both phase current and magnetic field waveform is observed (Fig. \ref{fig_TBHIF} (n,x)). 

\item Stable arcing, where distortion of both positive and negative half cycle occurs. The v-i plot has an 's' shape typical of the non-linear characteristic of HIF (Fig. \ref{fig_TBHIF} (t)).
\end{itemize}

For the low fault-current stages (initial stage and sizzling stage) the load current dominates, and the fault current characteristics cannot be readily observed in either the current or magnetic field measurements. A more detailed analysis of this data is given in \cite{Sifat2021}, where we have applied discrete Fourier Transform and Hilbert Transform algorithms to characterize the frequency and transeient characteristic of the TB-HIF stages.

\section{Evaluation of GMR Sensors for Detection of HIFs}
In summary, as shown in the example experimental results presented in section \ref{results}:
\begin{itemize}
\item There is no one set of HIF features. Different unique stages can occur in HIFs and their characteristics depend on the nature of the surface type which can also change as the fault progresses. These unique stages exhibit different fault current waveform shapes along with different visual and audible phenomena.

\item Arcing does not always occur in HIFs; low current non-arcing stages also occur.

\item Load current dominates over the low current non-arcing HIF fault behaviour.
These characteristics mean that HIF detection using simple algorithms is likely to be problematic and sophisticated pattern recognition algorithms will need to be applied to measurements in order to detect some HIFs. 
\end{itemize}

We have developed a classification algorithm, based on Deep Learning, to classify signals from the GMR sensors into 3 fault classes and one class of 'normal/no-fault' (Fig. \ref{fig_detectionAlg}). When tested on the experimental data from the fault test facility:

\begin{figure}
	\centering
	\includegraphics[width=0.5\textwidth]{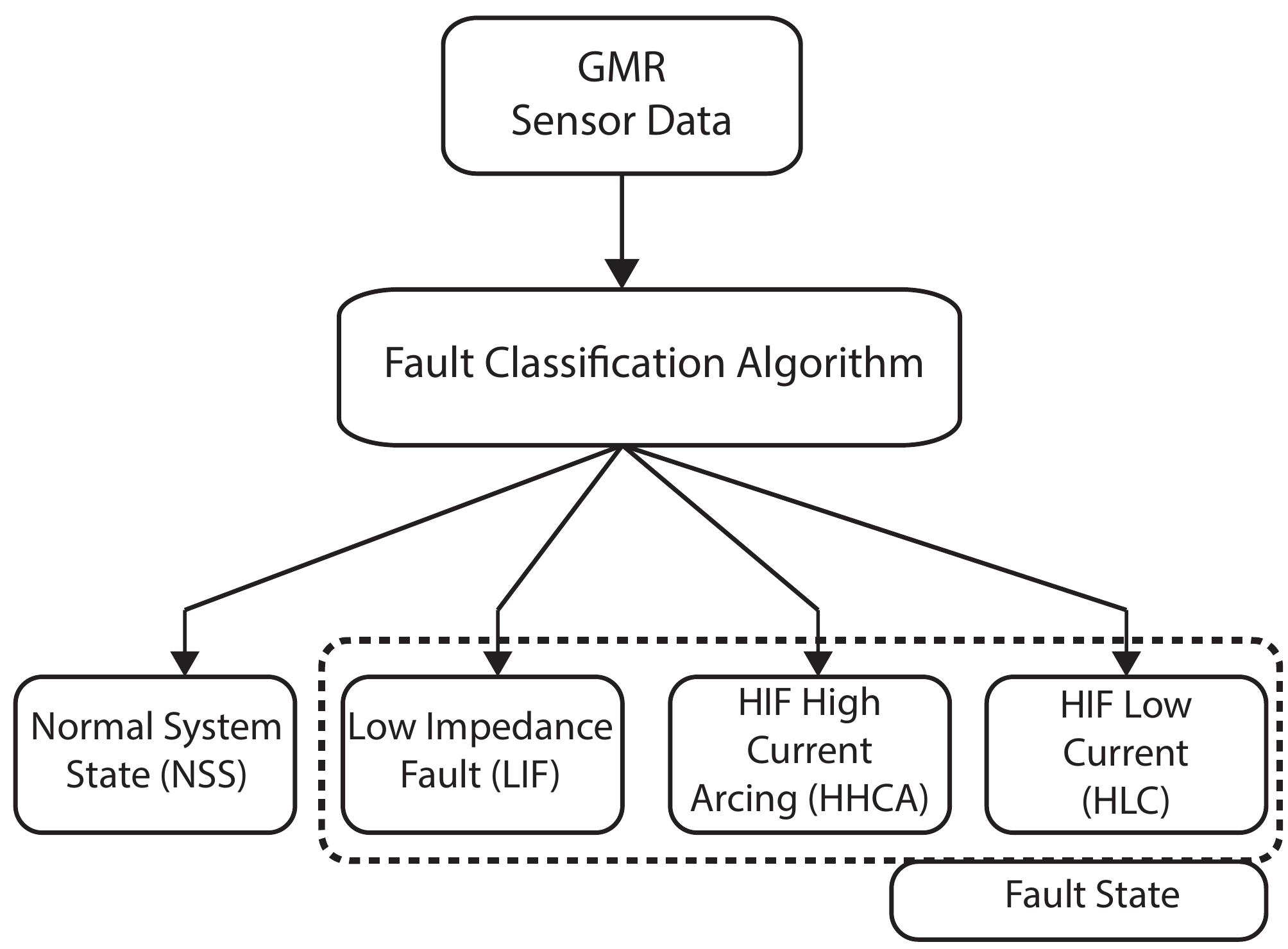}
	\caption{Fault pattern recognition framework.}
	\label{fig_detectionAlg}
\end{figure}

\begin{itemize}
\item 100\% of fault instances were classified as a fault of some type although 1.5\% of Arcing HIF were miss-classified as non-arcing HIF.
\item 99.98\% of normal system states were correctly classified as having no fault present.
\end{itemize}

The next step is to collect and test this pattern recognition system on data from real networks. Progress toward this is discussed in section \ref{outdoor}.

\section{Evaluation of GMR Sensor for Non-contact Current Sensing} 
\begin{figure}
	\vspace{-1em}
	\centering
	\includegraphics[width=0.85\textwidth]{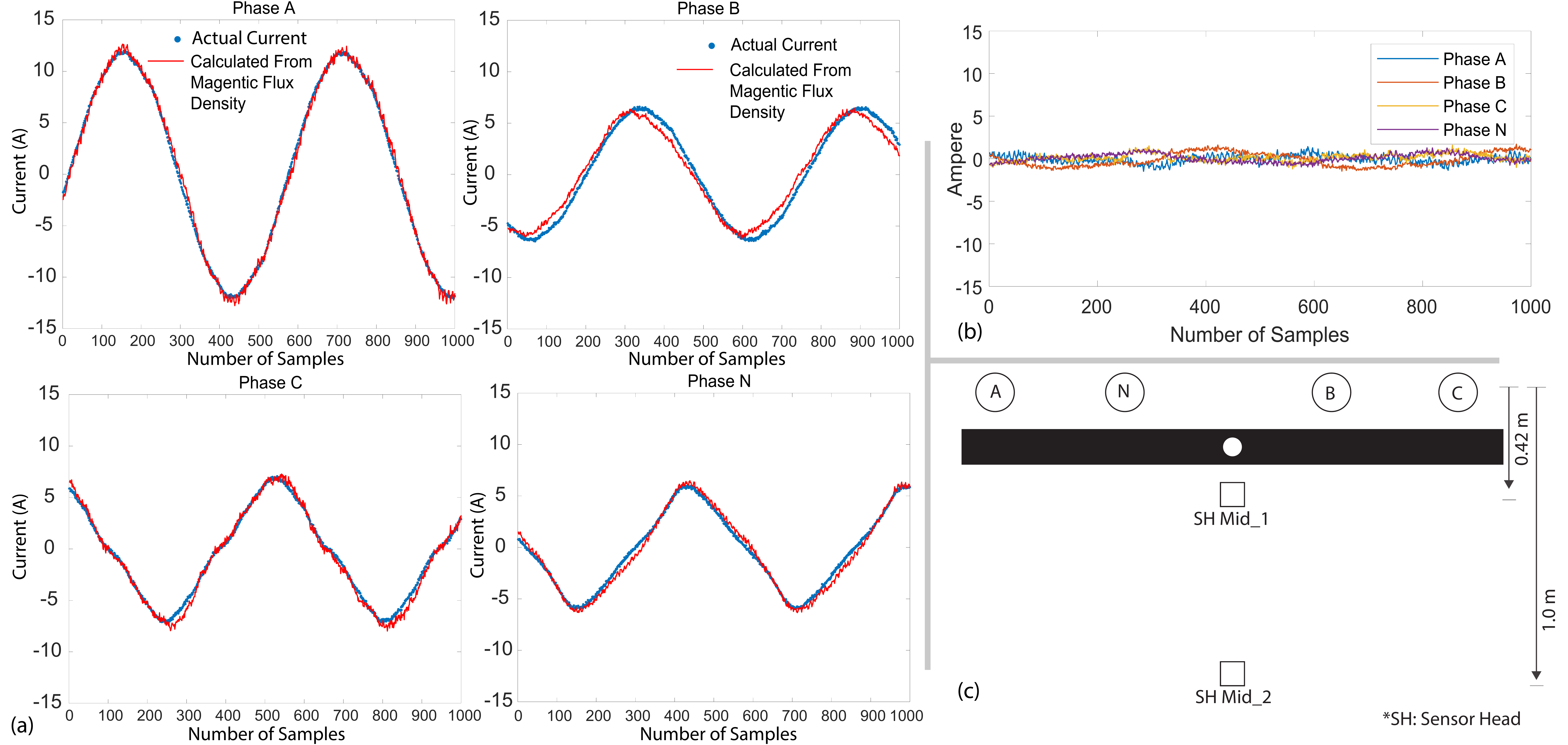}
	\caption[]{Comparison of load current measurement using two vertically placed sensors at 0.42m and 1m from the overhead lines (b) Instantaneous difference between actual and calculated current (c) Vertical combination of sensor heads placed 0.58 m apart.}
	\label{fig_backcalc}
	\vspace{-1em}
\end{figure}
As shown in the comparison of the GMR versus Hall Effect current measurements in section \ref{results}, the phase current behaviour is strongly reflected in measurements from a GMR sensor on the pole. As part of our project we have therefore also evaluated the use of GMR sensors as non-contact current sensors, in addition to their use for fault detection.

For non-contact current sensing we have evaluated the use of two 2D GMR sensor heads (x- and z-axes) vertically located on the pole. We mathematically decouple the overhead line current from the GMR sensors measurements. The comparison between the current measured with the Hall Effect sensors versus those calculated from the GMR sensor measurements shows a good match (Fig. \ref{fig_backcalc} (a)). The instantaneous current error was less than 1 A (Fig. \ref{fig_backcalc} (b)) and the root mean square (RMS) error between the actual and calculated currents ranged from 1.3-5.3\%.

\section{Pole Mounted Sensing System Prototype}\label{outdoor}
\begin{figure}
	\vspace{-1em}
	\centering
	\includegraphics[width=0.8\textwidth]{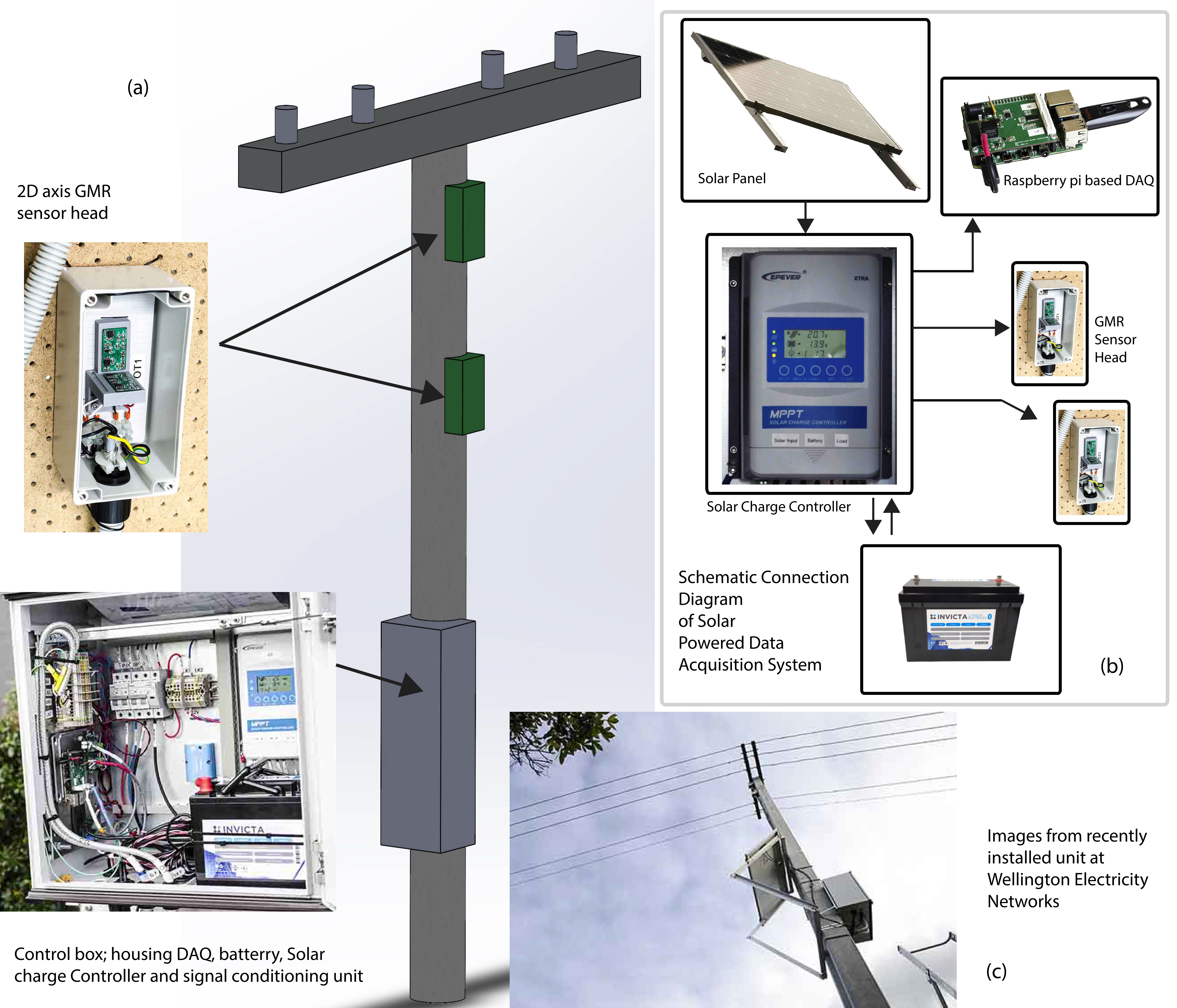}
	\caption[]{Outdoor GMR sensor based current sensing unit.(a) Schematic of electrical power pole and location of sensor head and control box. (b) Connections between solar power system for data acquisition and sensor head units. (c) Outdoor monitoring system installed in the Wellington Electricity overhead line networks.}
	\label{fig_outdoorinstl}
	\vspace{-1em}
\end{figure}

In parallel to our lab test facility experimental work, we have developed a prototype outdoor sensing system and have installed it in the Wellington Electricity (WE) network, near Gracefield, Lower Hutt substation (Fig. \ref{fig_outdoorinstl} (c)). The outdoor prototype includes GMR sensor heads, and a data acquisition and storage system. The system was designed as a standalone device, thus, it is powered using off the shelf battery pack, solar panel, and charge controller unit with maximum power point tracker (MPPT) capability (Fig. \ref{fig_outdoorinstl} (b)). With this we are collecting data from two different locations on the pole to observe the spatially cross-coupled fields from the four overhead lines (Fig. \ref{fig_outdoorinstl} (a)).

The data acquisition module was developed in-house using a embedded device, \textit{Raspberry Pi} integrated with an analog to digital converter (ADC). Data is sampled at 27.7 kHz and transferred to the raspberry pi for storage via serial peripheral interface (SPI) protocol (Fig. \ref{fig_outdoorinstl} (b)). The raspberry pi typically runs on Debian operating system, which provides a graphical user interface (GUI) to easily monitor and program the module on the fly. The module can also be accessed over the wireless fidelity (WIFI) protocol.

We are currently continuously monitoring and collecting data from this outdoor unit. We will use this data to confirm and refine our pattern recognition algorithms for fault detection and to refine our hardware selections.

\section{Conclusion}
In electricity distribution networks, fault detection is an ongoing challenge, particularly of High Impedance faults. GMR sensors are an attractive proposition for more widespread monitoring and as a basis for fault detection because of their broad bandwidth, low power consumption, miniature size, ability to digitally interface with intelligent systems and low cost. Our experiments in our purpose-built lab test facility have confirmed the dynamic response of these sensors during fault events and load changes. 

We have characterized the dynamic behaviour of fault currents and magnetic fields. A number of new and unique stages of HIF faults have been identified, dependent on surface material type and duration of the fault. FOr example as shown here, a number of new and unique stages of TB-HIF faults have been identified. Arcing does not always occur. It would be desirable to detect HIF faults in real networks even during the non-arcing stages, but they cannot be readily observed in raw current or magnetic field measurements. 

In our work we have found that suitable signal processing techniques can reveal the characteristic features of these non-arcing states and that Deep learning algorithms can automatically detect and classify fault states versus normal system events. Our current focus is to confirm and refine our algorithms using data collected in real networks, starting initially with our prototype outdoor sensing system installed in Wellington Electricity’s network.

\section*{Acknowledgment}
This work was financially supported in part by the New Zealand Science for Technological Innovation National Science Challenge (contract RTVU1702); Ministry of Business, Innovation and Employment New Zealand (contract RTVU1811) and Victoria University of Wellington Research Trust.
 
We would also like to acknowledge the support of Wellington Electricity, who have installed a prototype version of our hardware in their network for ongoing research data collection.

\ifCLASSOPTIONcaptionsoff
  \newpage
\fi

\bibliographystyle{IEEEtran}
\bibliography{thesref_1}

\end{document}